\documentclass[a4paper,11pt]{article}
\usepackage{pos}

\title{The limits of QGP-like effects towards smaller systems: from Pb-Pb down to pp and fixed-target collisions}
\ShortTitle{The limits of QGP-like effects towards smaller systems}

\author*[a]{Nicol\`o Jacazio}

\onbehalf{for the ALICE collaboration}

\affiliation[a]{University of Bologna and INFN,\\
  Via Irnerio 46, Bologna, Italy}

\emailAdd{nicolo.jacazio@cern.ch}

\abstract{
  Experimental findings of recent years blurred the frontier between large and small systems.
  The features attributed to the Quark Gluon Plasma formation have also been found in smaller systems when measuring particle production in high multiplicity events.
  These common features arise in multiple sectors, namely the particle dynamics (known as collective flow) and also when considering hadrochemistry (e.g., strangeness enhancement).
  The limit in small systems where this non-trivial behaviour occurs, is of very high interest in the field and is actively being investigated.
  This is carried out by performing multi-differential analyses and by selecting collision systems that are smaller than pp collisions.
  The current experimental contour of the limits between large and small systems is discussed in these proceedings.
}

\FullConference{The Eleventh Annual Conference on Large Hadron Collider Physics (LHCP2023)\\
  22-26 May 2023\\
  Belgrade, Serbia\\}

\begin{document}
\maketitle

\section{Introduction}

Quantum Chromodynamics (QCD) describes the nature of the strong force that drives the interactions between quarks and gluons.
These are the elementary particles that constitute protons, neutrons, and other hadrons.
By its nature QCD in the low energy regime can only be studied with a non-perturbative approach.
However in regimes of high energy density ($\epsilon > 1\ {\rm GeV/fm}^{3}$) the degrees of freedom of ordinary matter (hadrons) are replaced by those of its elementary components.
In this state of matter quarks and gluons are no longer confined within hadrons but are deconfined, with increased mean-free path.
These are the conditions reached in ultrarelativistic heavy-ion (AA) collisions, where the energy density is sufficiently high to produce the deconfined medium.
As such, the emergent features that were historically proposed as signatures of the QGP formation can be found in the final state resulting from a AA collision.
Such signatures have been observed in data from SPS, RHIC and LHC, giving a set of compelling evidence of QGP formation in ultrarelativistic AA collisions \cite{ALICE:2022wpn}.

In contrast to AA collisions, small systems (e.g. pp and and p--A collisions) were considered as a reference for the AA case, assuming that no deconfined phase could be produced.
They were used as a control experiment to disentangle the effects of cold nuclear matter in the collisions.
This representation was challenged in the LHC era.
In particular, several features of a collective, equilibrated medium, also expected as signatures of this deconfined QCD medium were observed in small systems.
In AA collisions the presence of long-range angular correlations between particles is expected to be due to the common origin of hadrons, in a collective evolution.
Results from CMS \cite{CMS:2010ifv} showed striking similarities between the two particle correlations measurements in Pb--Pb collisions and pp collisions.
This was obtained only when selecting events with a large number of charged particles produced.
Later, the same observation was made in p--Pb collisions \cite{CMS:2013jlh, ALICE:2012eyl, ALICE:2013snk}.

After the LHC measurements, the definition of small system can be revised starting from large systems.
This attempt defines a small system as a system being a priori too small to show characteristics of large systems physics, showing them nonetheless.
It has to be noted that in this representation there is no rigorous definition of a system too small to exhibit AA phenomenology.
In addition, using small systems as reference for the AA case is still valid as high multiplicity events constitute a small fraction of the total cross section in pp collisions.
Yet, the frontier of the QGP-like effects in small systems remains blurred.
The effort of the LHC experiments is being focused in better defining this border.
This is done by scanning small fractions of the cross section of the small collision systems and by performing multi-differential analyses that challenge models of particle production.

\section{Results and discussion}

\begin{figure}[h]
  \centering
  \includegraphics[width=0.4\textwidth]{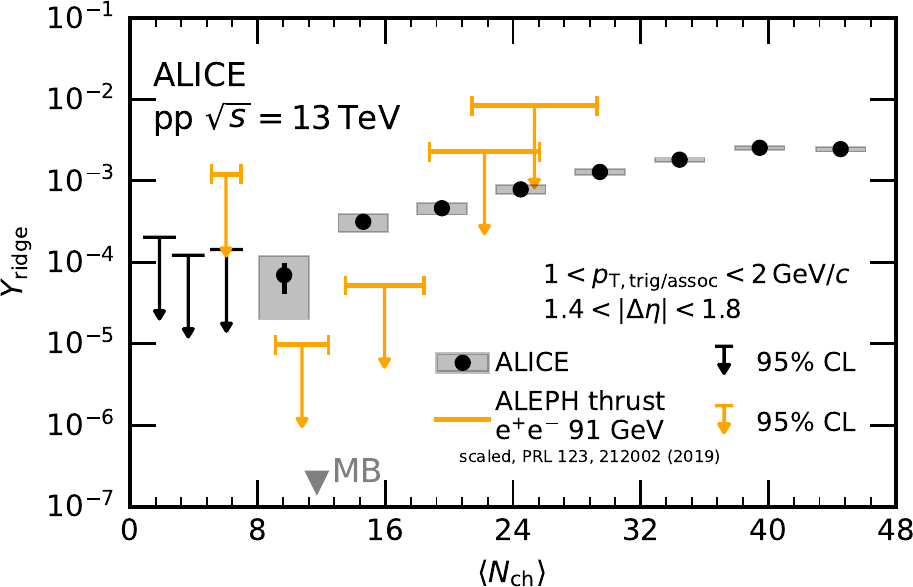}
  \caption{
  Ridge yield as a function of multiplicity, compared to the upper limits on the ridge yield in $e^{+}e^{-}$ collisions \cite{ALICE:2023ulm}.
  The orange limits represent the measurement in the thrust-axis reference frame with ALEPH with limits given at 95\% CL.
  }
  \label{lowmult}
\end{figure}

$e^{+}e^{-}$ collisions constitute the lower limit in size of small systems as defined by multiplicity, and the impinging particles are themselves pointlike.
Therefore they are ideal to explore the onset of the QGP-like effects
In this case, one does not find long-range correlations even when selecting high-multiplicity events \cite{Badea:2019vey}.
Comparing $e^{+}e^{-}$ and low-multiplicity pp collisions in Fig.~\ref{lowmult} one sees that the correlation is stronger in hadronic collisions at similar multiplicities.

\begin{figure}[h]
  \centering
  \includegraphics[width=0.39\textwidth]{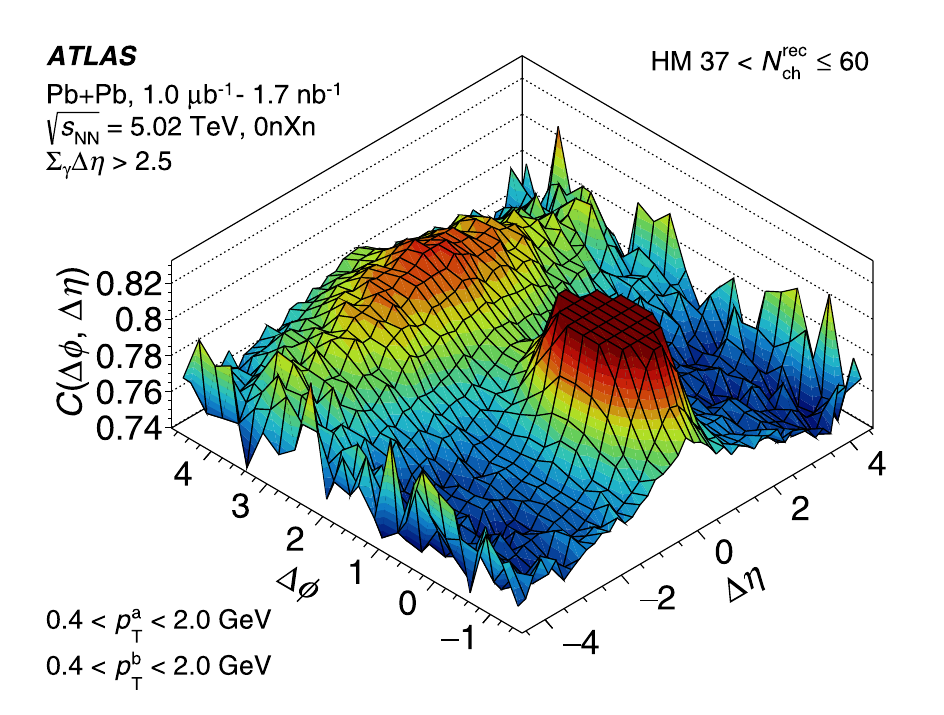}
  \includegraphics[width=0.39\textwidth]{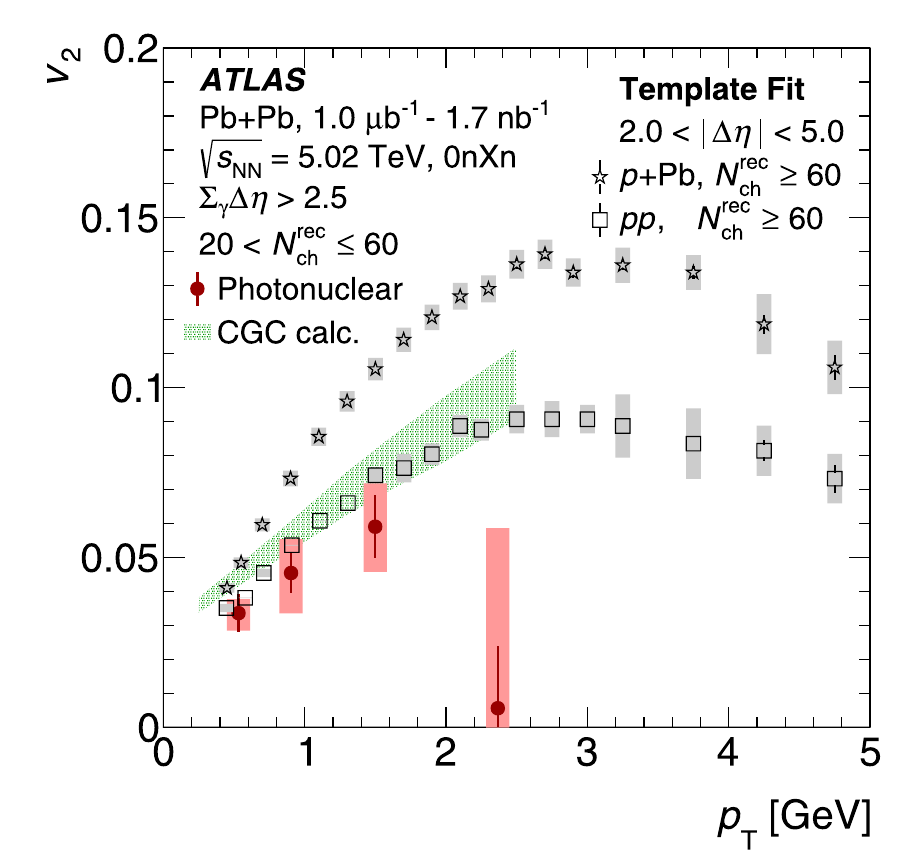}
  \caption{
    Two-dimensional normalized charged particle pair distributions in photonuclear high-multiplicity events as a function of $\eta$ and $\varphi$ and $v_{2}$ coefficients as a function of $p_{\rm T}$.
  }
  \label{upc}
\end{figure}

Another possibility to study small systems is to use ultra-peripheral collisions with ions.
Coulomb fields of relativistic moving charges are equivalent to a flux of photons boosted at high energies.
Photons can interact with the incoming protons or nucleus, obtaining at the LHC $\gamma {\rm p}$ and $\gamma {\rm Pb}$ collisions \cite{CMS:2022doq, ATLAS:2021jhn}.
At the LHC, the $\gamma$ energies are about tens of GeV with 2.5 TeV Pb beams.
Ultra-peripheral events with higher multiplicity show no clear near side ridge as reported in Fig.~\ref{upc}.
Non-zero elliptic flow of charged particles is found even in these events (red markers of right Fig.~\ref{upc}).
Measurements show weaker flow with respect to hadron-hadron collisions.
Nevertheless, one should consider that $v_{2}$ coefficients are vulnerable to non-flow effects.

\begin{figure}[h]
  \centering
  \includegraphics[width=0.4\textwidth]{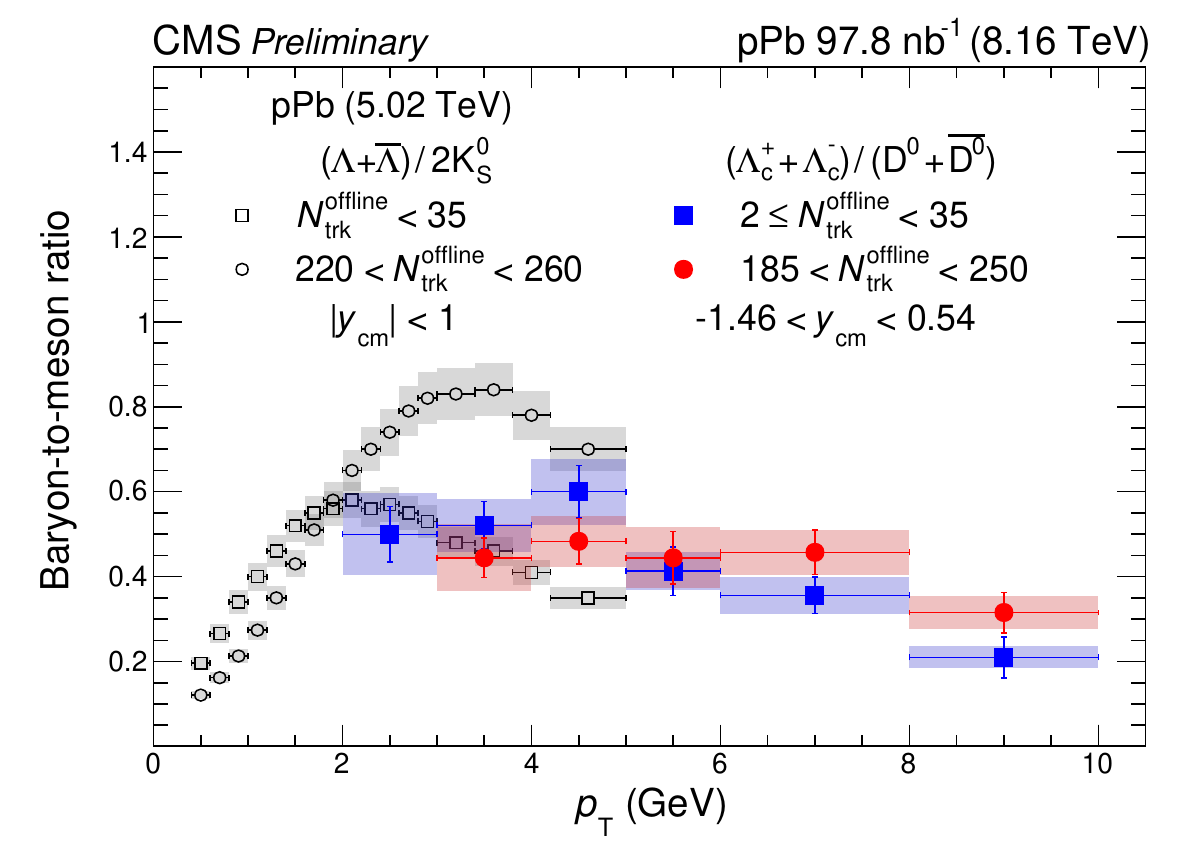}
  \includegraphics[width=0.4\textwidth]{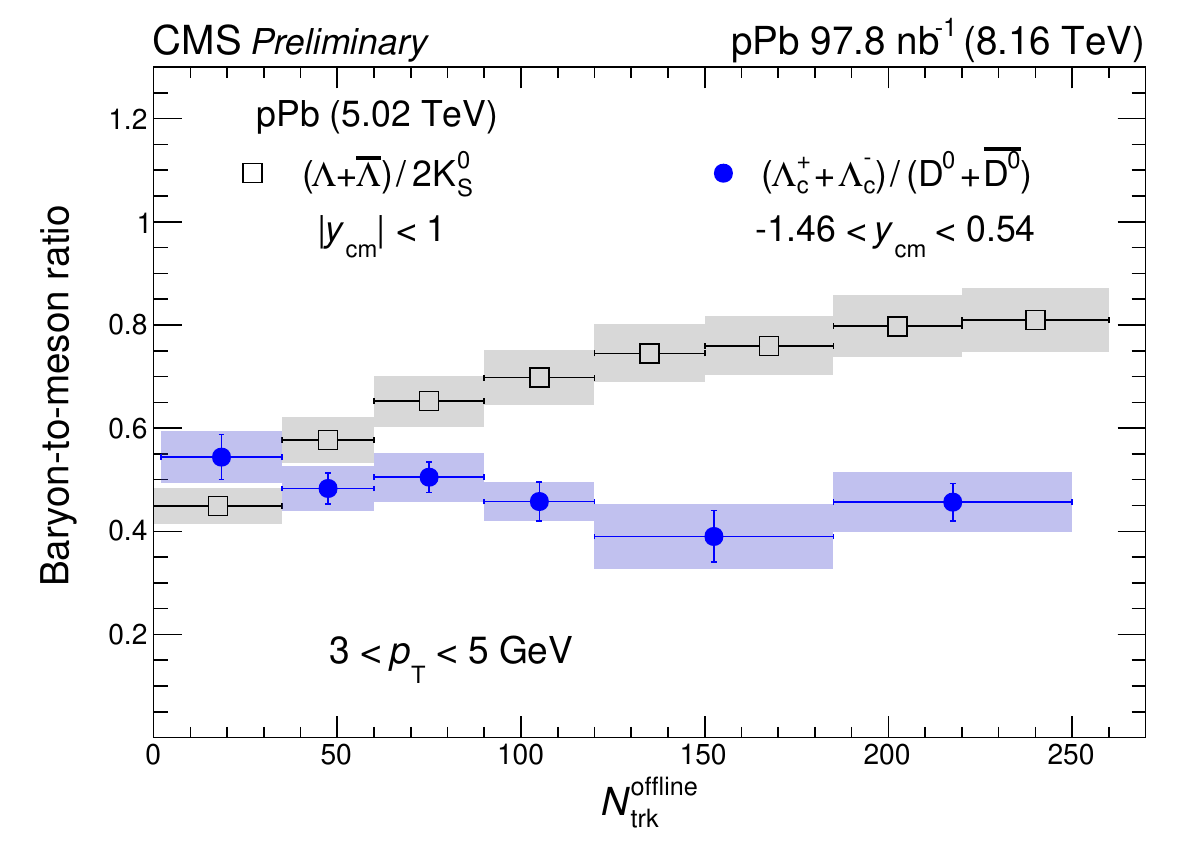}
  \caption{
    $\Lambda_{c}^{+} / D^{0}$ and $\Lambda/ K^{0}_{s}$ measured in p--Pb collisions as a function of $p_{\rm T}$ for high and low multiplicity (left) and multiplicity at intermediate $p_{\rm T}$ (right).
  }
  \label{baryon}
\end{figure}

A feature that distinguishes large and small systems can be found when comparing baryon and meson production.
This is particularly evident in the heavy flavour sector, e.g. it was found how $\Lambda_{c}^{+} / D^{0}$ is enhanced at intermediate $p_{\rm T}$ in central Pb--Pb collisions \cite{ALICE:2021bib}.
However in p--Pb, $\Lambda_{c}^{+} / D^{0}$ seems not to depend on the final state multiplicity \cite{CMS:2023oui} as shown in Fig.~\ref{baryon}.
In the light sector, instead, the $\Lambda/ K^{0}_{s}$ is seen increasing, indicating that the coalescence of heavy quarks might stop sooner than for light quarks in small systems.

Strangeness enhancement was historically proposed as a signature of the QGP formation in AA collisions, today it is generally seen as a sign of the system reaching equilibration.
It has been shown in recent years how this behaviour scales as a function of multiplicity independently of the collision system.
To pinpoint the origin of this effect, multidifferential measurements as a function of the charged particles multiplicity and effective energy (the energy available for the system to produce particle at mid-rapidity) are carried out in pp collisions.

\begin{figure}[h]
  \centering
  \includegraphics[width=0.8\textwidth]{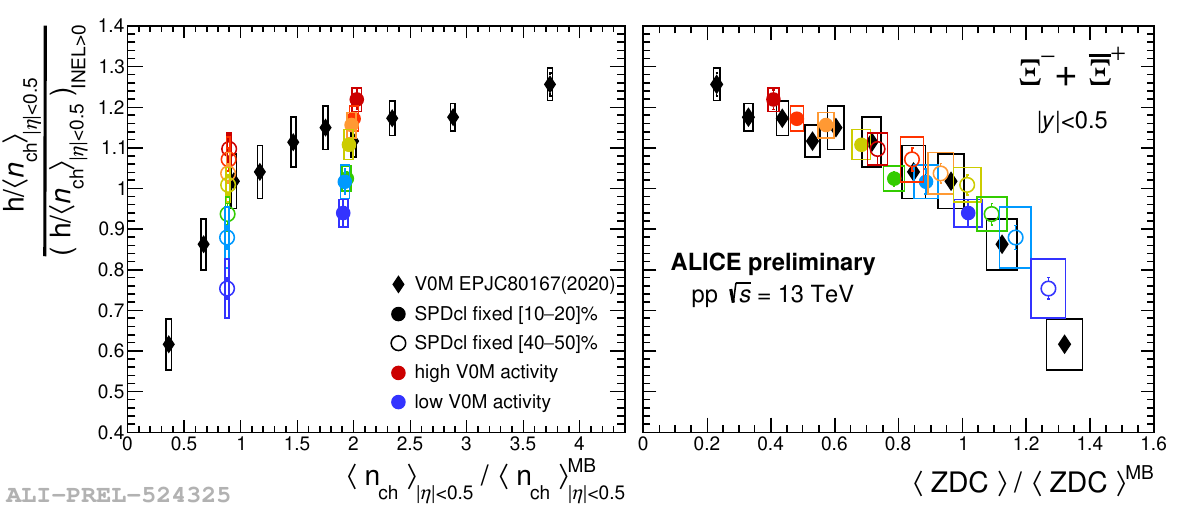}
  \caption{
    Ratio of the $\Xi^{-}+\overline{\Xi}^{+}$ to non strange hadrons normalized to the multiplicity integrated case as a function of the relative multiplicity and $\rm \left<ZDC\right>/\left<ZDC\right>^{MB}$, i.e. the forward energy, complementary to the effective energy.
  }
  \label{zdc}
\end{figure}

An increase in the average fraction of strange hadrons with increasing multiplicity and increasing effective energy is observed, as shown in Fig.~\ref{zdc}.
This modulation is also observed as a function of the effective energy when keeping the multiplicity fixed.

\begin{figure}[h]
  \centering
  \includegraphics[width=0.34\textwidth]{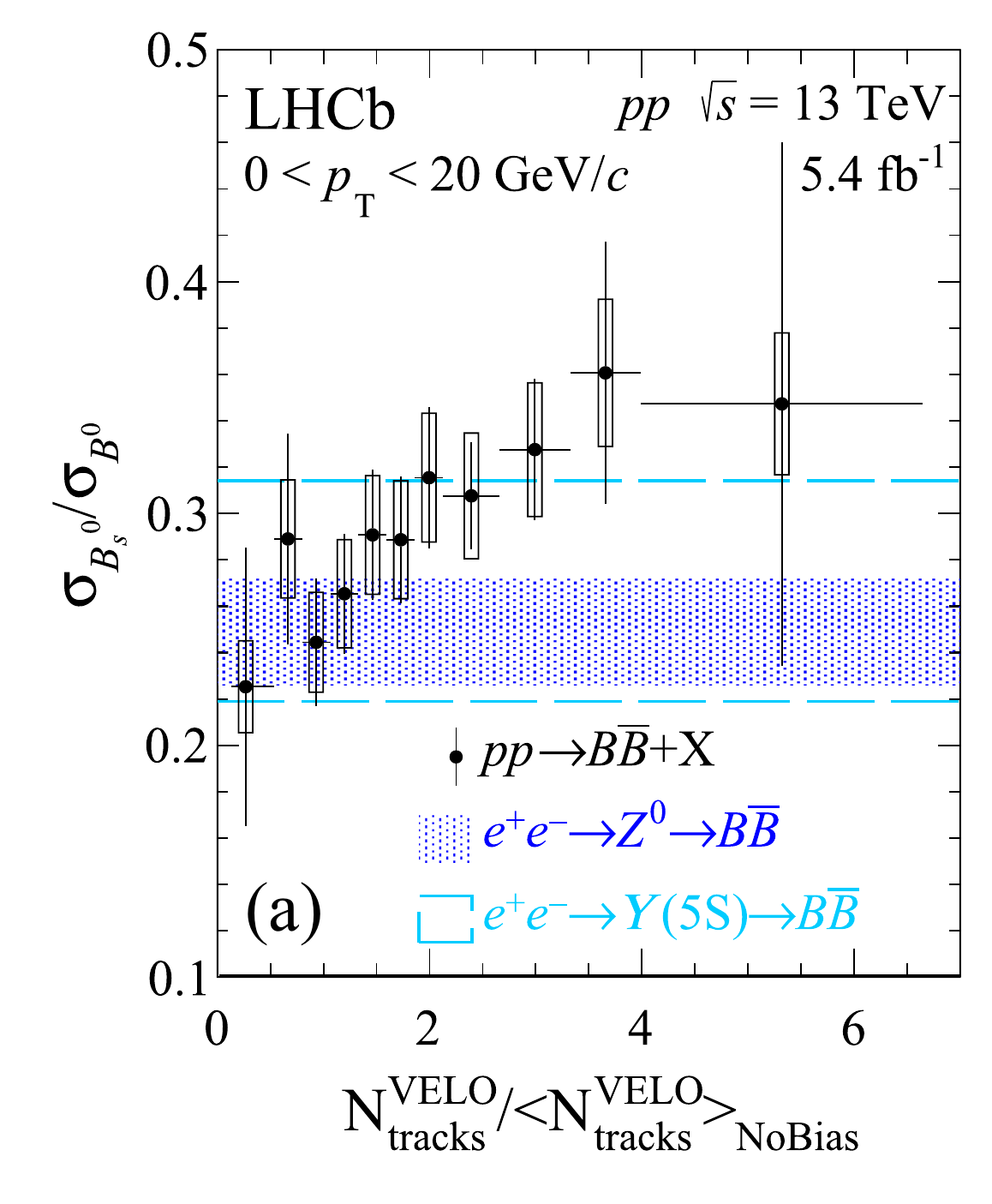}
  \caption{
  Ratio of cross sections $\sigma_{B^{0}_{s}}/\sigma_{B^{0}}$ versus the normalized multiplicity.
  The horizontal bands show the values measured in $e^{+}e^{-}$ collisions.
  }
  \label{beauty}
\end{figure}

The measurement of $B_{s}^{0}$ and $B^{0}$ as a function of multiplicity in pp collisions \cite{LHCb:2022syj} show a significant increase of $B_{s}^{0}/ B^{0}$ with multiplicity when measured in the same rapidity range as shown in Fig.~\ref{beauty}.
In this case, $b\bar{b}$ pair production is dominated by hard parton-parton interactions in the initial stage of the collision.
The enhancement might be due to quark coalescence, as more $s$ quarks are available at high-multiplicity the formation of $B_{s}^{0}$ is preferred over $B^{0}$.

\begin{figure}[h]
  \centering
  \includegraphics[width=0.4\textwidth]{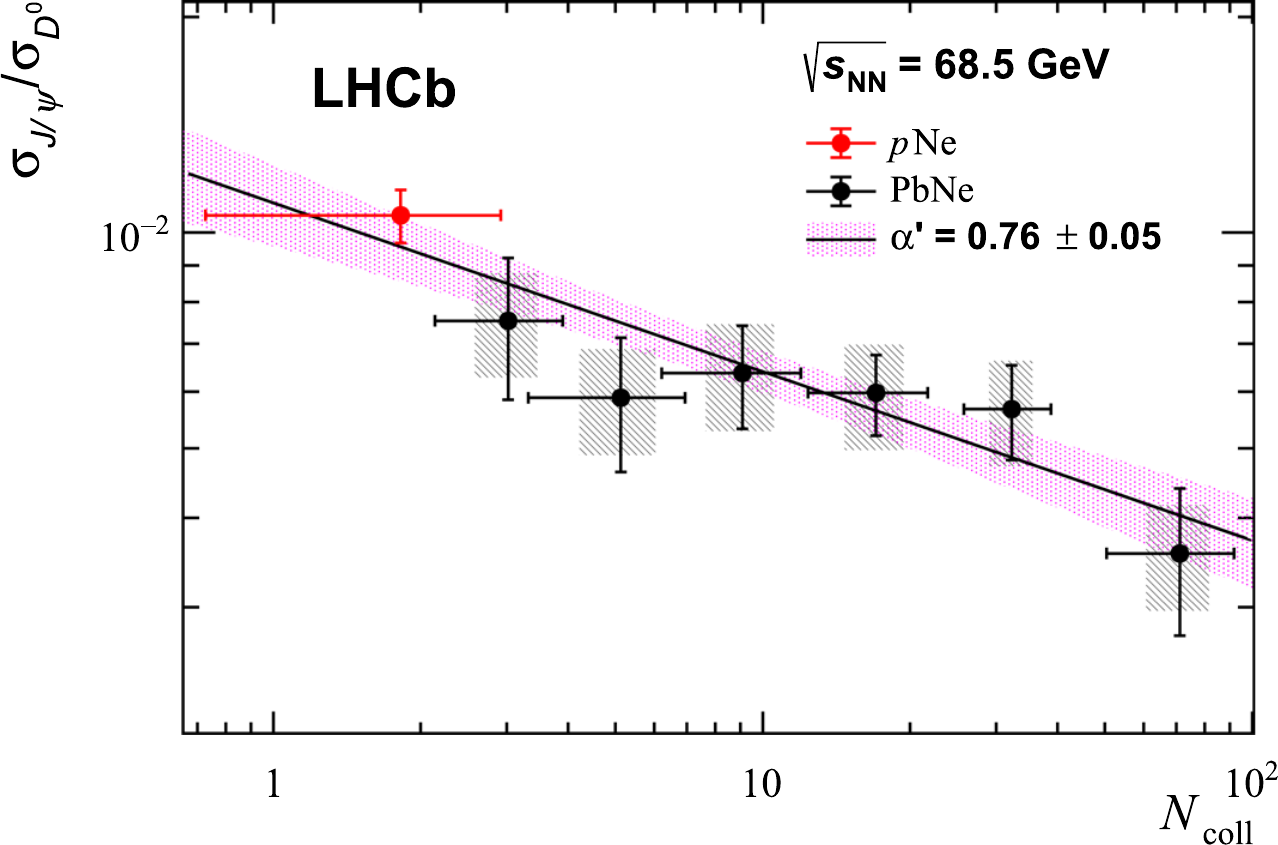}
  \caption{
    $\sigma_{J/\psi} / \sigma_{D_{0}}$ as a function of the number of binary collisions measured in p--Ne and Pb--Ne collisions.
  }
  \label{fixed}
\end{figure}

The fixed target program of the LHCb with the SMOG system can be used to scan small systems at the LHC bridging between pp, p--A and AA.
Measurements of $J/\psi$ production in p--Ne collisions \cite{LHCb:2022qvj} seem to indicate a universal behaviour when comparing p--Ne with Pb--Ne as a function of multiplicity the $\sigma_{J/\psi} / \sigma_{D_{0}}$, as shown in Fig.~\ref{fixed}.
In order to complete the picture, however, additional measurements are needed by varying collision system and increasing data samples.
In this respect, the upgraded SMOG2 system will allow for fixed target data taking in parallel to collider mode, resulting in an increased expected statistics of $J/\psi$ by more than three orders of magnitude \cite{Citron:2018lsq}, allowing also for more collision systems to be studied.

\section{Conclusions}
Thanks to the LHC measurements in recent years, the border between small and large collision systems has become blurred, with the former exhibiting features typical of the latter.
The difference between large and small collision systems is being investigated by performing multi-differential analyses and searching systems too small to show AA phenomenology.
It was shown that $e^{+}e^{-}$ collisions do not show clear signs of collective behaviour even at higher final state multiplicities.
Yet, in photonuclear collisions, non-zero flow is observed even if no long-range correlations are found on the near side.
Baryons and mesons are also showing different behaviour between small and large systems.
While $\Lambda_{c}^{+}$ is enhanced in central Pb--Pb events, there is no evident enhancement in p--Pb collisions in the heavy flavour sector, possibly related to the way that heavy flavour quarks are produced in the initial partonic scatterings.
Studies as a function of multiplicity and effective energy highlight how the latter plays a role in the observed enhancement of strange hadrons.
However, results on $B$ hadrons in small systems as a function of the multiplicity show how the abundance of $s$ quarks in high multiplicity collisions might enhance mesons with strange content if quark coalescence is at play.
For the future, the usage of large samples with more collision systems seems promising.
Results with smaller collision systems in the fixed target configuration highlight no significant discrepancies between p--Ne and Pb--Ne in the $J/\psi$ production.
This frontier will be significantly expanded by extending the statistics and available collision systems with upgraded detectors.


\end{document}